\documentclass[twocolumn]{aastex701}
\renewcommand{\cite}{ \citep}

\usepackage{amsmath}
\usepackage{graphicx}
\usepackage{subcaption}

\usepackage[utf8]{inputenc}
\usepackage[T1]{fontenc}
\usepackage{mathptmx}
\usepackage{siunitx} %
\usepackage{hyperref}
\usepackage{microtype}

\begin{document}

\title[Pair-Alfvén Shocks]{Formation and Evolution of Pair-Alfvén Shocks with PIC Simulations}

\author{J. D. Riordan} 
\affiliation{Department of Physics, University College Cork, Cork T12 K8AF, Ireland}
\email{dannyriordan@gmail.com}

\author{M. E. Dieckmann} 
\affiliation{Department of Science and Technology, Link\"oping University, SE-60174 Norrk\"oping, Sweden}
\email{mark.e.dieckmann@liu.se}

\author{A. Pe'er} 
\affiliation{Department of Physics, Bar-Ilan University, Ramat-Gan, 5290002 Israel}
\email{peerasa@biu.ac.il}

\date{\today}

\begin{abstract}
    We consider the recently discovered \emph{pair-Alfvén shock wave} occurring in collisionless
	electron-positron plasmas. We perform a series of Particle-In-Cell studies in one and two
	dimensions in order to determine the stability conditions for such a shock and
	the mechanisms which sustain its growth. Building on our previous
	simulations, which established that these shocks are initially mediated by the
	Weibel instability before becoming Alfvénic, we demonstrate that the shock is
	sustained by self-generated Alfvén waves overtaking the shock in the upstream plasma.
	As a result growth is only possible when the guiding magnetic field strength, and hence Alfvén
	speed, is sufficiently small, $\omega_c \lesssim 0.4$ (in normalized units).
	Furthermore the production of the waves in the upstream is dependent on a
	resonance between the Alfvén wave mode and the thermal noise in the plasma,
	which is inhibited at high magnetization. This explains the conditional absence of this
	type of shock identified previously.

\end{abstract}

\section{Introduction}

Relativistic jets are known to exist in a great variety of distinct astronomical objects. They are composed of electrons, positrons and some unknown fraction of ions. These pairs may already exist at the launching of the jet or be produced by photon-photon pair annihilation in the extreme conditions that exist near the jet base, close to a central compact object (e.g. black hole or neutron star)\cite{JoachimBlome1988,Bottcher1999,Dieckmann2019a}. This hot, rarefied
plasma is often well approximated as being \emph{collisionless}, i.e. the rate of direct
particle-particle interactions is negligible, and the plasma dynamics are
mediated primarily through collective interaction with both externally applied and
self-generated $B$-fields. Within this scenario, fluctuations in the flow will lead to internal shocks, which will dissipate kinetic energy and lead to fluctuations in observed lightcurves\cite{Rees94}.

The so-called \emph{pair-Alfvén shock}, as described in our previous work\cite{Dieckmann2020}, is a newly discovered type of shock defined by the following: (1) it occurs in plasmas primarily composed of electrons and positrons; (2) it is initially launched by a filamentation instability as the upstream plasma flows into a slower moving medium; and (3) it is sustained in the steady state by Alfvén waves which are generated downstream and build up at the shock discontinuity.

Plasma instability mechanisms which cause flow of energy from the fluid motion into plasma waves, and determining which instability dominates in a given environment is an area of an ongoing study. Contingent on the plasma conditions (e.g. densities of the colliding plasmas) there are many relevant  instabilities that can lead to the formation of collisionless shocks \cite{Marcowith16,Bret2018}.
In the case of a shocked pair plasma as considered in this work, typically the fastest growing mode will be the filamentation instability, however other modes, such as oblique modes which occur when the net magnetic field is not aligned with the shock normal, can initially outgrow the filamentation mode and preheat the upstream plasma\cite{Dieckmann2019}. 

The two-stream instability creates a thermal anisotropy, with higher temperature along the collision direction. %
Furthermore, it is possible that particles, which are emanated by the shock into the upstream region, trigger the growth of additional instabilities, such as the Bell instability \cite{Bell2004, Riquelme2009}.

In our previous work\cite{Dieckmann2020} we identified and characterised this kind of shocks in the case of a weak magnetic field. Here, we extend this by examining more strongly magnetized environments to determine the requisite conditions for the shock's formation and stability.
In particular we consider the conditions under which the  shock forms and
examine the role of instabilities resulting from the interaction of shock-heated
plasma countering-streaming in the inflowing material.

This work is therefore significant in improving our understanding of trans-relativistic, collisionless, pair-plasma environments, such as those predicted to occur at e.g. at the base of relativistic jets in black hole microquasars\cite{Pelletier01}, active galactic nuclei\cite{Bottcher1999}, or compact binary millisecond pulsars\cite{Linares21}. In particular shocks in these environments provide potential sites for efficient first-order Fermi processes \cite{Sironi2015} and hence are of great interest for the investigation of cosmic ray production\cite{Sironi09}.

Astrophysical plasma phenomena of this kind are typically studied using Particle-In-Cell (PIC) simulations. This is because PIC simulations, while computationally expensive, do not rely on heuristic models of the plasma behaviour, and allow recovering information on the particles and electromagnetic fields directly. We do however, out of computational necessity, make the approximation of using one- and two-dimensional spatial grids. We assert\cite{Dieckmann2020,Stark17} that these reduced-dimensionality simulations still capture the essential physics and that the results are still qualitatively applicable to real-world 3D scenarios.

Despite this there are notable limitations to the reduced-dimensionality simulations. Some notable drawbacks are: in 1-dimensional simulations the filamentation instability is inhibited due to the alignment of the simulation box with the magnetic field. As a result, only the two-stream instability contributes to the shock growth and hence the shock takes an artificially long time to form. In the 2D geometry filamentation modes with in-plane magnetic fields are excluded, implying again a slowing of pair-Alfvén shock generation.

This paper is organized as follows. In Section 2 we discuss the theory of
pair-Alfvén shocks. In Section 3 we discuss the simulation details. Our
results are presented in Section 4. We discuss our findings and conclusions in Section 5.

\section{Background}

\subsection{Pair-Alfvén waves in a thermal-equilibrium plasma}

Throughout this work we use naturalised units in order to simplify presentation. Frequencies are normalized to the total pair plasma frequency
$\omega_p=\sqrt{2}\omega_{pe}$, where, $\omega_{pe}$ is the plasma frequency of
electrons with density $n$. Time is therefore given in units $\omega_p^{-1}$. We normalize distance to the skin depth $\lambda_s = c / \omega_p$. 
Electric fields $\mathbf{E}$ are normalized to $m_e c \omega_p / e$ and magnetic
fields $\mathbf{B}$ to $m_e \omega_p / e$, where $c$ is the speed of light, $e$ is the electron charge and $m_e$ its mass.

The dispersion relation for a wave in a warm, magnetized pair plasma was derived by \citet{Keppens2019} (their equation 4.2). This equation, when written in natural units, reads 
\begin{align}
\begin{split}
    \label{eqn:kgd19hotdispersionOMA}
       \omega^{6} 
     - \omega^{4} \left(B^{2} + k^{2} v_s^{2} + k^{2} + 2\right)  \\
     - \omega^{2} \left(- B^{2} k^{2} \Lambda^{2} v_s^{2} - B^{2} k^{2} - B^{2}  \right.\\
     \phantom{\omega^2 } \left. - k^{4} v_s^{2} - k^{2} v_s^{2} - k^{2} - 1\right) \\
     - B^{2} k^{4} \Lambda^{2} v_s^{2} - B^{2} k^{2} \Lambda^{2} = 0. 
\end{split}
\end{align}
Here, $\omega$ is wave frequency, $B$ is the magnetic field strength, $\mathbf{k}$ is the wavevector, $k=|\mathbf{k}|$ is the wavenumber, $v_s$ is plasma sound speed, and the wave orientation relative to the magnetic field is measured by $\Lambda = {\mathbf k \cdot \mathbf B}/{kB}$. The sound speed $v_s$ can be taken as $0$ for a cold plasma.

 We assume the presence of a "background" magnetic field $\mathbf{B}_0$ with the amplitude $B_0 =\left|\mathbf{B}_0\right|$, and we orient our coordinate system such that it is aligned with the $x$ axis. Components of particle velocity and wavevectors parallel and perpendicular to this field will be denoted $v_\parallel, k_\parallel$ and $v_\perp, k_\perp$ respectively.

Although in general there exist six different wave solutions, by taking $\mathbf{k} \parallel \mathbf{B}_0$ and hence $\Lambda=1$ we recover the familiar Alfvén solution, 
$$\omega=\Lambda kB/\sqrt{1+B^2}=v_Ak,$$
where $v_A=B/\sqrt\mu$ is the non-relativistic Alfv\'en speed and $\mu$ is the charge density.
We focus on the low-frequency regime in which only this pair-Alfvén wave exists.

In the limit of low $k$, cold pair Alfvén waves have phase
speed $v_\phi = \omega / k = B_0 / {(2\mu_0 n m_e)}^{1/2}$, where $\mu_0$ is the vacuum permeability.  The charged particles gyrate at the normalized electron gyro-frequency, $\omega_c = eB_0/\omega_p m_e$. In the limit of high wavenumber $k$ and $\omega_c < 1$ the wave frequency converges to $\omega_c$, the cyclotron resonance \cite{Keppens2019}. 

Particles will resonate with  pair-Alfvén waves if they meet the resonance condition: $\omega = \omega_c \pm \mathbf{v}\cdot\mathbf{k}$. In our simulations, the highest energy particles drawn from the thermal distribution have a speed approximately $4 v_t$ (where $v_t = \sqrt{k_B T_e /m_e}$ is the thermal speed, and $T_e$ is the electron temperature) and this determines the range of velocity angles for which resonant interaction can occur.

\subsection{Two-stream and firehose instability}
\label{two-stream-and-firehose}

As established in our previous work \cite{Dieckmann2020}, structures are seen to form in the
density distributions corresponding to electrostatic ``holes''  in phase space, i.e. localized areas with relatively low or zero density. The observed structure in the 
downstream region results when these phase space holes collapse and disappear. While the holes can be stable in 1D simulations, they eventually collapse in higher dimensions. These collapses heat the plasma and thus serve to transfer energy from the counterstreaming beams into isotropic thermal motion. In that work we found the holes are present until $t\approx 500$ (here and throughout references to times in the simulation are given in our normalized units) when they are disrupted by the growth of the pair-Alfvén waves.

\begin{figure*}
	\includegraphics[width=\textwidth]{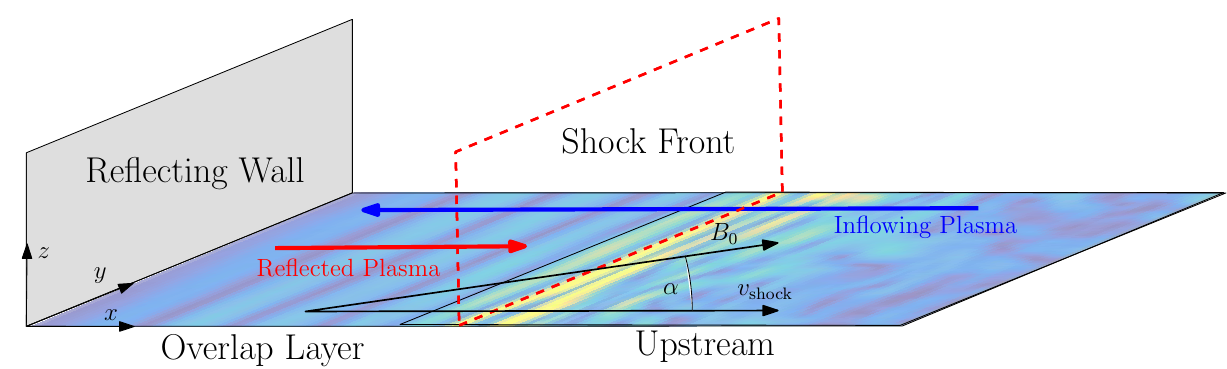}
	\caption{{
    Sketch of simulation domain. We distinguish the following sub-regions: \emph{overlap
	layer} -  where the inflowing plasma and the plasma reflected from the
    left wall meet upstream, here the plasma is cold in the fluid frame;
    \emph{downstream} - expanding from $x=0$, this region contains the plasma which
    has been processed by the shock and is dense and hot in all directions;
    \emph{foreshock} - the region where energetic particles have escaped from
    downstream and the overlap area and interact with the inflowing upstream plasma. 
    Note the the $z$ dimension is not resolved, and that in the 2D simulations where the $y$ dimension is resolved we have used a periodic boundary condition.
    }
    }
\label{shocksketch}
 \end{figure*}

We note two distinct regions of the plasma with counterstreaming beams which
give rise to a two-stream instability (see sketch in \autoref{shocksketch}). Firstly in the overlap layer we have
beams flowing into and out of the shock with almost identical properties.
In the upstream we have a dense beam as well as a shock-heated pair component.
Hence we expect to find two distinct scenarios for pair-Alfvén instability
growth corresponding to each of these regions.

Firehose-type instabilities are produced at parallel shocks when a threshold
number of particles with $v_{\parallel} \approx \omega / k_\parallel$ is reached and an
anisotropic velocity distribution is present\cite{Bret2020,Bret2022,Bret2023}. It will be useful to define the parallel and perpendicular temperatures, $T_\parallel = T_x$ and $T_\perp=T_y + T_z$. These refer to the temperature in two orthogonal projections of the full 3-dimensional spatial domain, the first a line parallel to the shock normal and the other a plane parallel to the shock plane, such that the total temperature is given by $T=T_\perp + T_\parallel$ . The anisotropy condition is  (for the downstream) $T_\perp / T_\parallel < 1 - 1/\beta$ where $\beta=2\mu_0nk_BT/B^2$ is the ratio of kinetic to magnetic pressure, and $k_B$ is the Boltzmann constant. We can restate this in terms of the ratio of Alfv\'en ($v_A$) and thermal speeds, 

\begin{equation}
   \label{anisotropy-vs-alfven-speed}
   \frac{T_\perp}{T_\parallel} < 1 - \frac{\mu_0}{2} \frac{v_A^2}{v_t^2}. 
\end{equation} 

Since these firehose modes depend on the presence of a thermal anisotropy, they will be quenched if the plasma distribution returned to isotropy, for example by a separate instability. As we will show, this can occur in the downstream of the environments considered in this work where the two-stream instability will thermalize the plasma along $x$. The above analysis cannot be applied to the foreshock, since it cannot be well-approximated by a single distribution with different parallel and perpendicular thermal speeds. In this case a fully kinetic treatment is possible \cite{Pokhotelov06} and the anisotropy condition $T_\perp \neq T_\parallel$ is replaced by $$\left(\omega-k_\parallel v_\parallel\right)\frac{\partial f}{v_\perp\partial v_\perp} \neq k_\parallel \frac{\partial f}{\partial v_\parallel},$$ where $f$ is the momentum distribution function.

\section{Simulations}
\subsection{Theory}

We use the plasma physics simulation code EPOCH \cite{Arber2015}. This code
numerically solves Maxwell's equations on a fixed grid to time-evolve a
relativistic system of electromagnetic waves and charged particles. For this
task, it uses an explicit second-order scheme, the Particle in Cell (PIC) method
\cite{Birdsall1985}. In PIC method, collections of physical particles are represented
using a smaller number of “pseudoparticles”, and the fields generated by their
motion are calculated using a leapfrog finite-difference time-domain (FDTD)
method. The forces on the pseudoparticles due to the calculated fields are then
used to update the pseudoparticle position and momentum using the Boris Pusher
scheme \cite{Arber2015}. EPOCH is capable of accurately simulating a large variety of
plasma systems at the microphysical scale \cite{Arber2015}.

Amp\'ere's law $\mu_0 \epsilon_0 \dot{\mathbf{E}}=\nabla \times \mathbf{B}-\mu_0
\mathbf{J}$ and Faraday's law $\dot{\mathbf{B}}=-\nabla \times \mathbf{E}$ are
approximated on a numerical grid, where $\mathbf{E}$, $\mathbf{B}$ and
$\mathbf{J}$ are the electric field, the magnetic field and the current density, and $\epsilon_0$ is the vacuum permittivity. Each
plasma species $j$ is approximated by an ensemble of computational particles
(CPs). 
The $i^{th}$ CP has a charge-to-mass ratio $q_i/m_i$ that must match
that of the species $j$ it represents. The electromagnetic fields are coupled to
the CPs and the CPs are coupled to $\mathbf{J}$ via suitable numerical schemes
as implemented in the EPOCH code we use \cite{Arber2015}.

\subsection{Numerical Setup}

Our simulation begins with an isotropic electron-positron plasma, both species have densities $n_0/2$ and temperature $T_0 = 10~{\rm keV}$, which gives thermal velocity $v_t = 0.14c$.  

In the 1D simulations we resolve a box length $L_N = 2800$ (in simulation units normalized to $\lambda_s$) by $2\times 10^4$
simulation cells and each particle species by 100 particles per cell. The cell
size is therefore equal to the Debye length $\Lambda_D=v_t/\omega_p =0.14$ and smaller than the skin depth by a factor of $\approx 7$. The code runs until the simulation time $t_\text{sim} = 4\times 10^4$ is reached, which was observed to be long enough to allow for transient effects.
To ensure that the box size is sufficiently large, we further perform an additional 1D
simulation with the same plasma parameters, using $10^7$ CPs for each species. The simulation box length is $L_x = 21270$, which we resolve by 148000 cells. The grid cell size is then $\Lambda_D \approx \lambda_s/7$ as before.

Our two-dimensional simulations resolve $x$ by $2\times 10^4$ grid cells and $y$
by $2000$ grid cells. Boundary conditions are reflective along $x$ and periodic
along $y$. We model one electron species and one positron species, which are
uniformly distributed in space. Each species has a density $n_0/2$ and simulated by a total of $8 \times 10^8$ CPs. 

The simulation box spans the intervals $0 \le x \le 2650$ and $0 \le y \le 265$ which are likewise in simulation units normalized to $\lambda_s$. 
Both species have a Maxwellian velocity distribution with temperature $T_0
= 10$~keV, which gives the thermal speed $v_t = \SI{4.2e7}{\meter/\second}$. The bulk fluid speed is $v_b = -3v_t$ along $x$ (which corresponds to $0.42\textrm{c}$ in the simulation). 
The Debye length is $\Lambda_D=0.14$. %
A magnetic field $\mathbf{B}_0=(B_0,0,0)$ is present at $t=0$. $B_0$ is determined by the choice of the electron gyro-frequency parameter $\omega_c = eB_0/m_e\omega_p$ (numerically, both $B_0$ and $\omega_c$ are identical when expressed in the code's normalized units). The pair
Alfvén speed $v_A = B_0 / {(\mu_0 n_0 m_e)}^{1/2}$ is $0.7v_t$ ($0.1c$).
The 2D simulations are stopped at $t_\text{sim}=2500$ (in simulation units). We checked that this time is sufficiently long to capture all transient effects of interest.

In total we perform four simulations of a parallel shock in 1D (varying $\omega_c$), three in 1D for oblique shocks (varying the angle between the shock and the background field $\vartheta$), and finally four 2D simulations varying $\omega_c$ as in the first set. In this way we are able to observe what effects the limitations of the (computationally simpler) 1D simulations have on the shock dynamics, and understand how the shock obliquity affects the results. Finally we fulfill our primary goal of extending our prior work\cite{Dieckmann2020} to higher magnetic fields and describing the different regimes of pair-Alfv\'en shock behaviour.

\begin{table}
\begin{ruledtabular}
\begin{tabular}{lcr}
Cell        size        &  $    d_x             $  &            $\SI{8.000000e+01}{ m           }$   \tabularnewline
Total       electrons   &  $    N_e             $  &            $\SI{10000000    }{             }$   \tabularnewline
Number      density     &  $    n_0             $  &            $\SI{1.668744e+08}{ /m^{-3}     }$   \tabularnewline
Guide       B-field     &  $    B_0             $  &            $\SI{4.483000e-07}{ T           }$   \tabularnewline
Alfvén      speed       &  $    v_A             $  &            $\SI{3.243579e+07}{ m/s         }$   \tabularnewline
Plasma      frequency   &  $    \omega_p        $  &            $\SI{7.287631e+05}{ Hz          }$   \tabularnewline
Skin        depth       &  $    \lambda_s       $  &            $\SI{4.113716e+02}{ m           }$   \tabularnewline
Wave        scale       &  $    k_0             $  &            $\SI{2.430892e-03}{ m^{-1}      }$   \tabularnewline
RMS         momentum    &  $    p_0             $  &            $\SI{1.207073e-22}{ kg.     m/s }$   \tabularnewline
Drift       momentum    &  $    p_d             $  &            $\SI{1.145069e-22}{ kg.     m/s }$   \tabularnewline
Drift Lorentz factor    &  $    \gamma_d        $  &            $\SI{1.084348e+00}{             }$   \tabularnewline
Drift       velocity    &  $    v_d             $  &            $\SI{1.159242e+08}{ m/s         }$   \tabularnewline
Initial     temp.       &  $    T_0             $  &            $\SI{1.160000e+08}{ K           }$   \tabularnewline
Drift       momentum    &  $    p_\text{drift}  $  &            $\SI{1.145069e-22}{ kg.     m/s }$   \tabularnewline
Thermal     velocity    &  $    v_0             $  &            $\SI{5.929815e+07}{ m/s         }$   \tabularnewline
Gyroradius  (for $B_0$))&  $    r_g             $  &            $\SI{1.063570e+03}{ m           }$   \tabularnewline
\end{tabular}
\end{ruledtabular}
\caption{\label{figure1}Representative values of plasma parameters in the simulated system.}
\end{table}

\section{Results}

\subsection{Initial shock formation}

\begin{figure*}
	\includegraphics[width=\textwidth]{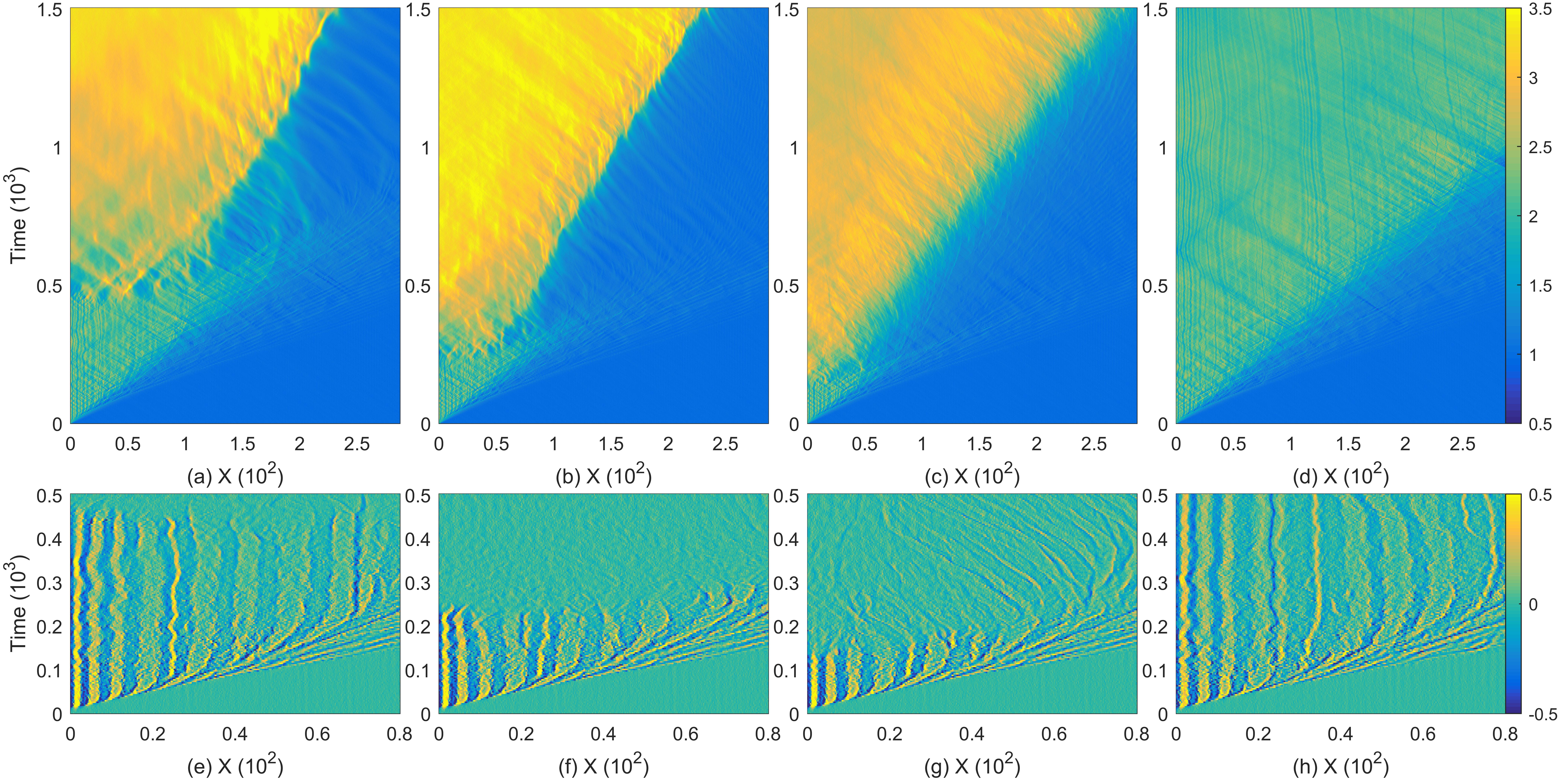}
	\caption{Initial evolution of the plasma in and near the overlap layer, showing the number density (upper panels) and charge separation (lower panels) as functions of space (x-axis) and time (y-axis). Panel (a)
	shows the electron number density $n_{e^-}$ in the simulation with $\omega_c =
	0.1$, (b) that in the simulation with $\omega_c = 0.2$, (c) that in the
	simulation with $\omega_c = 0.4$ and (d) that in the simulation with
	$\omega_c = 0.8$. Panels (e-h) show the difference $n_{e^+} - n_{e^-}$ between the
	number densities of positrons and electrons corresponding to the cases
	above. Panel (e) shows the difference for $\omega_c = 0.1$, panel (f) that
	for $\omega_c = 0.2$, panel (g) that for $\omega_c = 0.4$ and panel (h) that
	for $\omega_c = 0.8$.}
\label{initial_evolution}
 \end{figure*}

In \autoref{initial_evolution} we plot the early-time results of our 1D simulation in order to examine the initial shock formation process for four values of the electron gyro-frequency: $\omega_c = 0.1, 0.2, 0.4$ \text{and} $0.8$. From the figure, one sees a clear qualitative distinction in the shock formation processes.
In the $\omega_c=0.1,0.2$ cases (left panels), a distinct initial period during which the counterstreaming plasma has not yet developed is observed, until approximately $t=5\times10^2,~ 3\times 10^2$ respectively. 
During this phase, the shock grows as pair-Alfv\'en waves accumulate at the shock front (which we will see in \autoref{merged-1d-phase-space}), and a charge separation is apparent in the lower panels. Once this process is complete the shock will have reached a steady state and efficiently thermalizes the downstream leaving a homogeneous plasma behind.

In the third column ($\omega_c=0.4$) one observes a similar behaviour, however the downstream plasma retains some minor charge separation and the shock front is less sharp. Finally in the $\omega_c=0.8$ case, the formation of the shock is suppressed and no clear shock is observed to form.

This can be explained by considering the tighter constraints on shock stability imposed by the stronger magnetic field. In Ref.\cite{Bret2018}, it is shown that the range of thermal anisotropy values which are stable with respect to the mirror instability shrinks with increasing plasma $\beta$ as 
\begin{align}\label{thermal-anisotropy-bound}
 1-1/\beta \leq T_\perp/T_\parallel  \leq 1+1/\beta, 
\end{align}
 and hence the range of anisotropies behaves like $\propto \omega_c^{-2}$. Specifically in the $\omega_c=0.8$ case, and with the initial parameters from our simulation \autoref{thermal-anisotropy-bound} predicts a maximum allowed deviation of only $\left| T_\perp / T_\parallel -1\right| \lesssim 0.07$.

As discussed in our prior work \cite{Dieckmann2019a}, $T_\perp/T_\parallel  \neq 1$ may arise as a result of small initial anisotropies. Plasma-wave interactions including mirror and firehose instabilities will tend to increase any existing anisotropy, and may outcompete the thermalizing effect of the electromagentic fields at the shock boundary.
\subsection{Wave spectra}

\begin{figure*}[ht]
	\includegraphics[width=\textwidth]{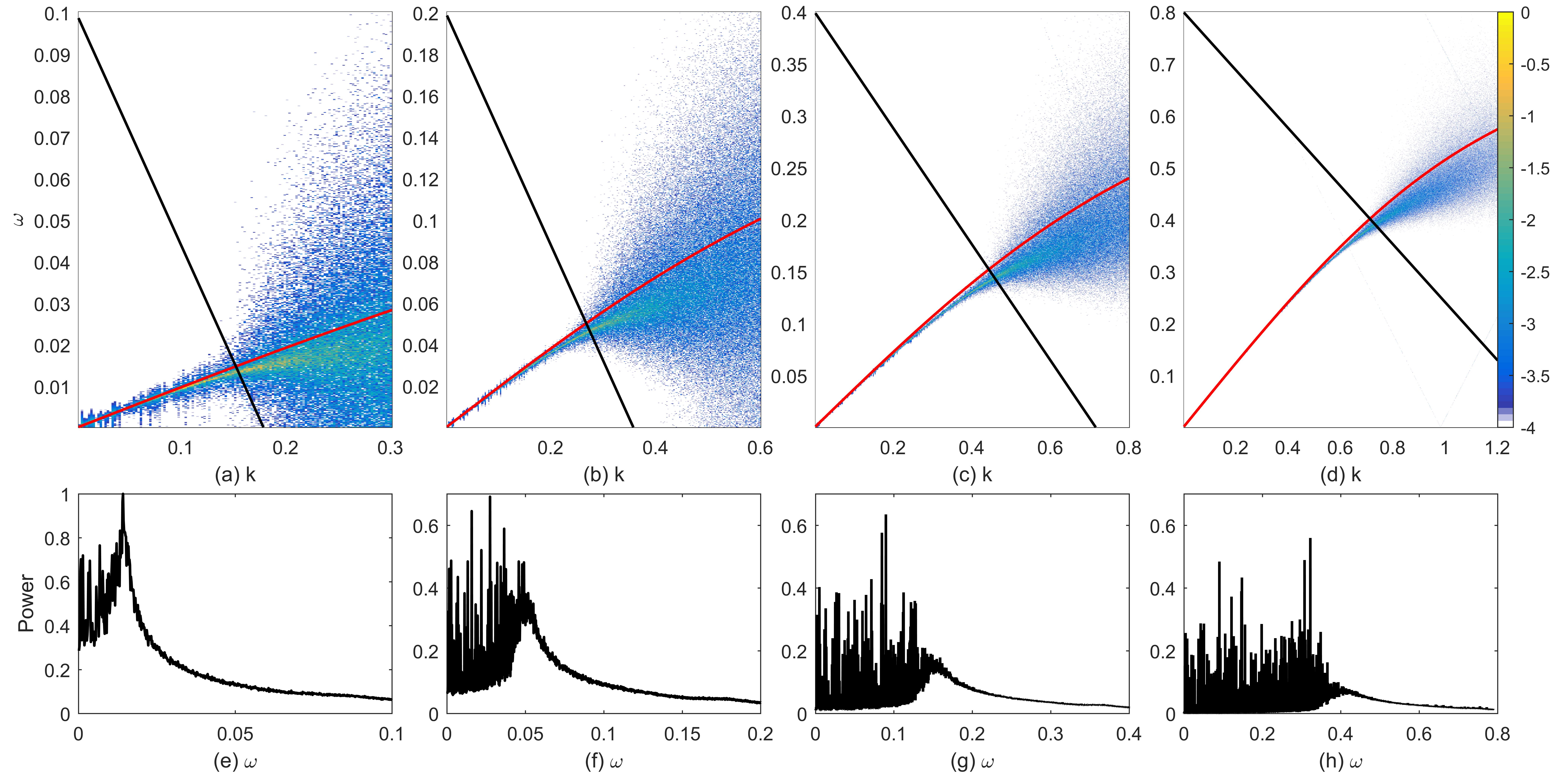}
	\caption{Power spectrum of $B$-field in 1D simulations for wave vectors parallel to the background magnetic field $\mathbf{B}_0$, along with the dispersion relation of pair-Alfvén mode (red) and line separating areas of low and high damping (black). Panels (a) and (e) correspond to the case $\omega_c = 0.1$. Panels (b) and (f) correspond to $\omega_c=0.2$. Panels (c) and (g) correspond to $\omega_c=0.4$. Panels (d) and (h) correspond to $\omega_c =0.8$}. 
	\label{dispersion_relation}
	\end{figure*}

In order to examine the energy distribution in the Alfve\'nic modes, we take the unshocked uniform plasma and Fourier transform the magnetic field energy density $B^2\left(x,t\right)$, as recorded over the full spatial and temporal range of the simulation, to $k-\omega$ space. The logarithm is plotted in \autoref{dispersion_relation}(a-d), normalized to the respective maxima.

We overplot two lines, the red line gives the analytic solution for pair-Alfvén wave with parallel magnetic field\cite{Keppens2019}.
The black line shows $\omega = \omega_c - v_t k$, which is the threshold for resonant interaction with the charged particles and hence delimits the region where the waves will be damped and deviate from the ideal wave solution. 
We find a sharp wave branch only to the left of this line, demonstrating good agreement with the expected Alfvén solution. Waves to the right of the black lines, while still following the analytical dispersion relation, interact strongly with particles and become increasingly damped as $k$ is increased (i.e., shorter wavelengths).

We obtain the power spectral density by integrating over all $k$ at each $\omega$ and display the results in ~\autoref{dispersion_relation}(e-h). When increasing the magnetic field, and hence cyclotron frequency $\omega_c$ (from left to right in \autoref{dispersion_relation}), the dispersion relation is modified, and the red lines become steeper. 
This results in a broadening of the observed spectrum (see panels (e) to (h)) and hence a broader range of velocities of Alfv\'en wave propagation. This weakens the condition of matching the Alfv\'en speed to the shock speed since waves at higher $k$ will be slowed.
The noise power peaks at the frequency where the waves start interacting with the particles and decreases steadily as we increase the frequency to $\omega_c$.

\begin{figure*}[ht]
	\includegraphics[width=\textwidth]{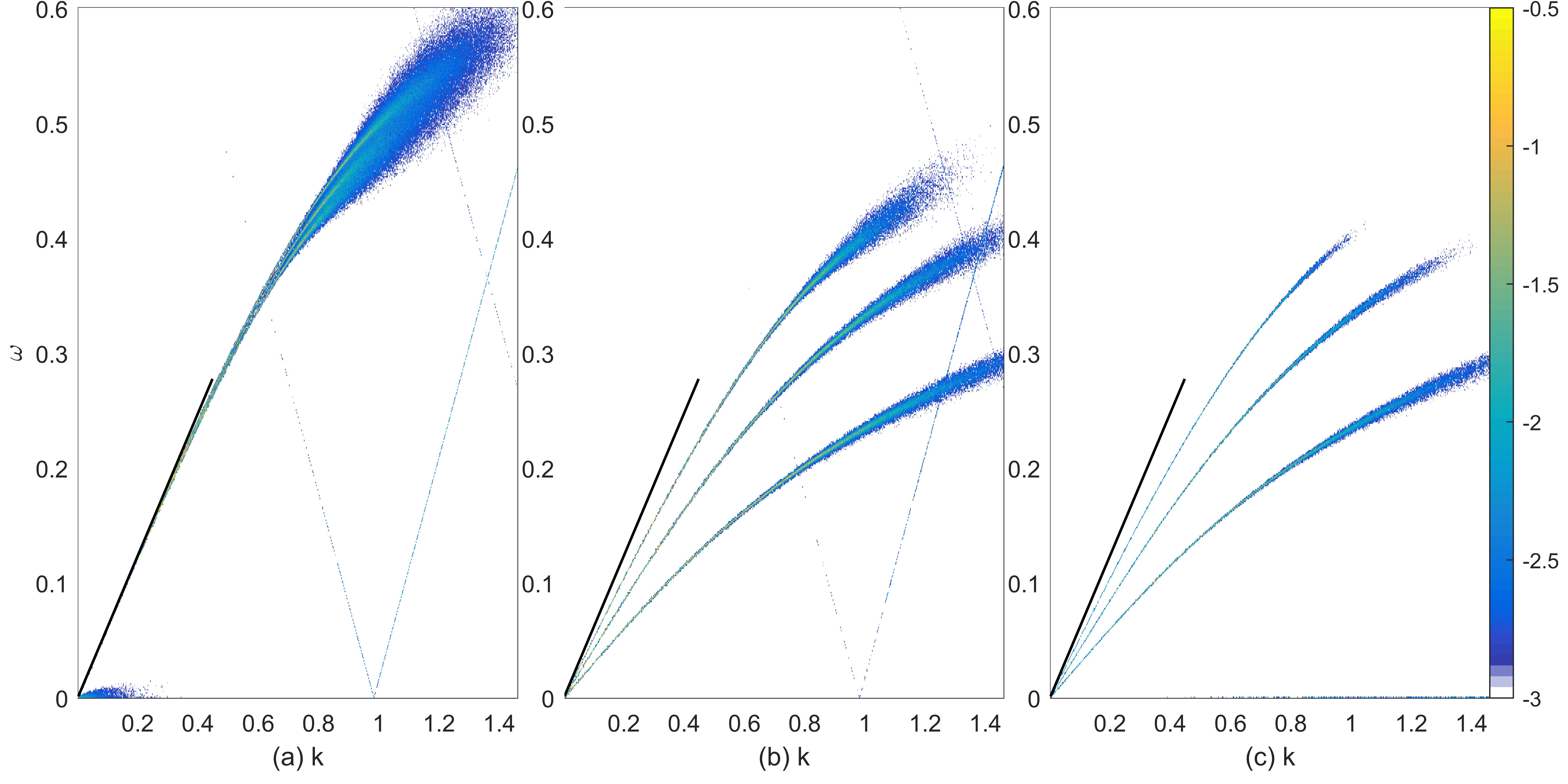}
    \caption{\
    Panels (a), (b), and (c) show the power spectrum of waves in $B_y$, $B_z$, and $E_x$ respectively. We have merged the results for obliquity angles $\alpha = 30^\circ$, $45^\circ$ and $60^\circ$ (relative to the $x$ axis). The branches overlap in (a), and can be identified in (b) and (c) by noting that increasing $\alpha$ results in a decrease in the slope near the origin (i.e. the phase velocity).
    Panel (a) shows $\log_{10}{\left(B_y^2/B_\textrm{max}^2\right)}$, (b) shows $\log_{10}{\left(B_z^2/B_\textrm{max}^2\right)}$
	and (c) shows $\log_{10}{\left(E_x^2/c^2B_\textrm{max}^2\right)}$, where $B_\textrm{max}^2=\max{B_y^2}$ is the peak of the power in the $y$-component of the $B$-field. The overplotted black line
	corresponds to the dispersion relation $\omega = v_A$ of the $\alpha=0$ Alfv\'en wave.
	The diagonal structures with low power are high-frequency $\omega \gg
	\omega_p$ modes that have been aliased due to the limited data sampling	rate. 
    }
	\label{oblique}
\end{figure*}

This setup also allows us to study the wave properties for the oblique case. We
perform simulations taking the same parameters as above but rotating the magnetic
field with strength $\omega_c = 0.8$ by each of $30^\circ$, $45^\circ$ and $60^\circ$ in
the $xy$ plane. The results are plotted in \autoref{oblique}. The power has been normalized to the peak value for each of the three simulations and all three power spectra have been shown together to allow for direct comparison. 

The waves shown in  \autoref{oblique} (a), which correspond to the power in $B_y$, change the amplitude of the background magnetic
field; we identify them as fast magnetosonic waves. As such, their propagation does not depend on a background magnetic field parallel to the wavevector, as is the case for Alfv\'enic modes. Their phase speed matches the Alfvén speed for $k<0.4$ and it hardly changes with the
propagation angle in this $k$-interval. Note that these modes do not exist when $\alpha = 0 $ (wave vector parallel to $\mathbf{B}_0$).

In contrast, for the waves in \autoref{oblique} (b) we observe the expected $\cos\alpha$ decrease in wave speed, as well as increased deviation from the expected result at higher wavenumber in the case of higher obliquity. These modes have a magnetic field component orthogonal to $\mathbf{k}$ and
$\mathbf{B}_0$ and their phase speed at low $k$ is $v_A \cos{\alpha}$. We identify them as oblique pair-Alfvén waves. 
These waves may have an electrostatic component, as is the case for our chosen plasma parameters (See \autoref{oblique} panel (c)). This figure confirms that the majority of the power in electromagnetic waves is carried in the normal and oblique Alfvén waves, the latter having both a magnetic and an electrostatic component.  

\subsection{Long-term evolution of the 1D shock}

\begin{figure}[h!]
    \begin{subfigure}{0.475\textwidth}
    \includegraphics[width=\textwidth]{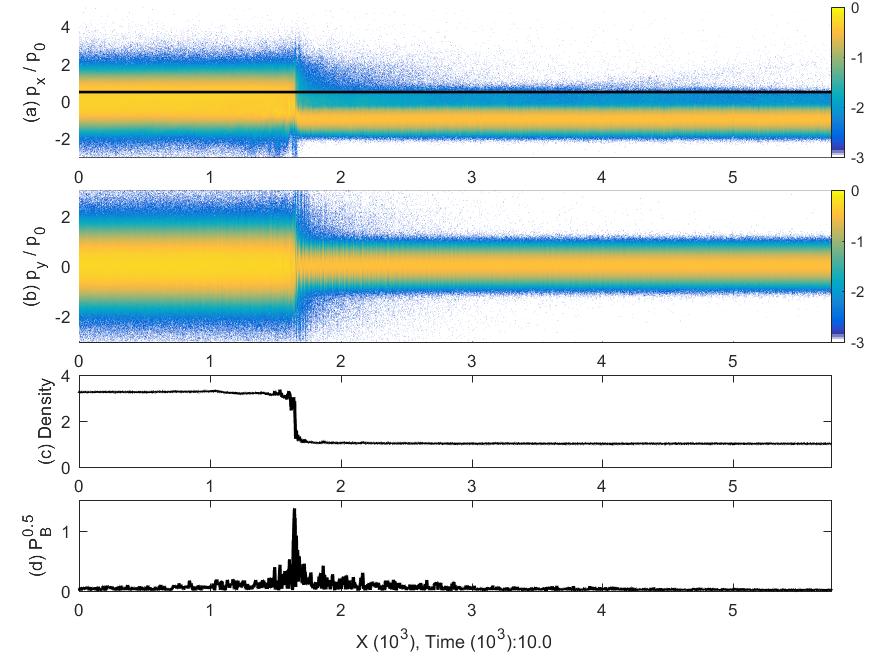}
	\caption{$\omega_c =0.2$}
    \label{w0.2-1d-phase-space}
    \end{subfigure}
    
    \begin{subfigure}{0.475\textwidth}
	\includegraphics[width=\textwidth]{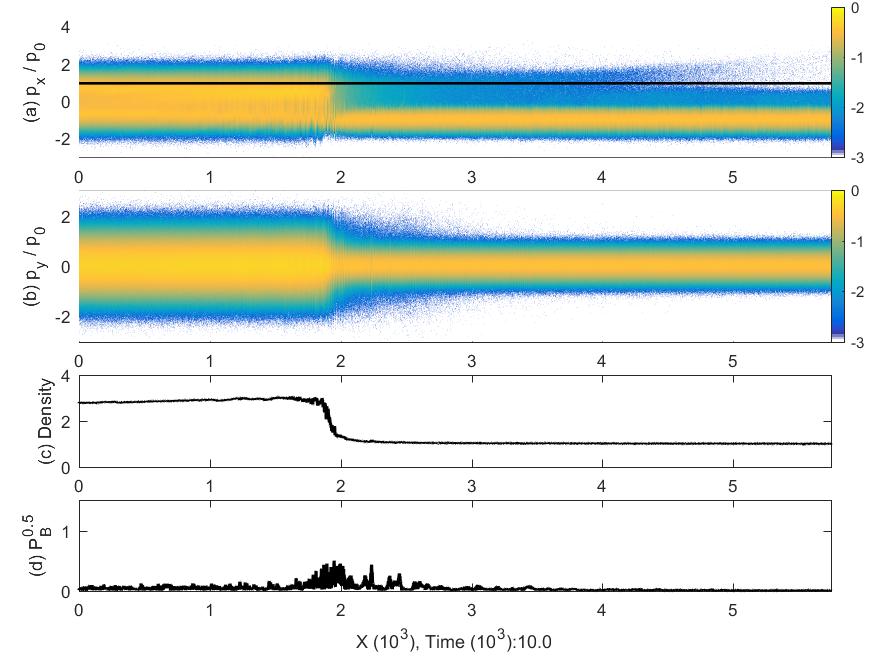}
	\caption{$\omega_c =0.4$}
    \label{w0.4-1d-phase-space}
    \end{subfigure}
    \begin{subfigure}{0.475\textwidth}
	\includegraphics[width=\textwidth]{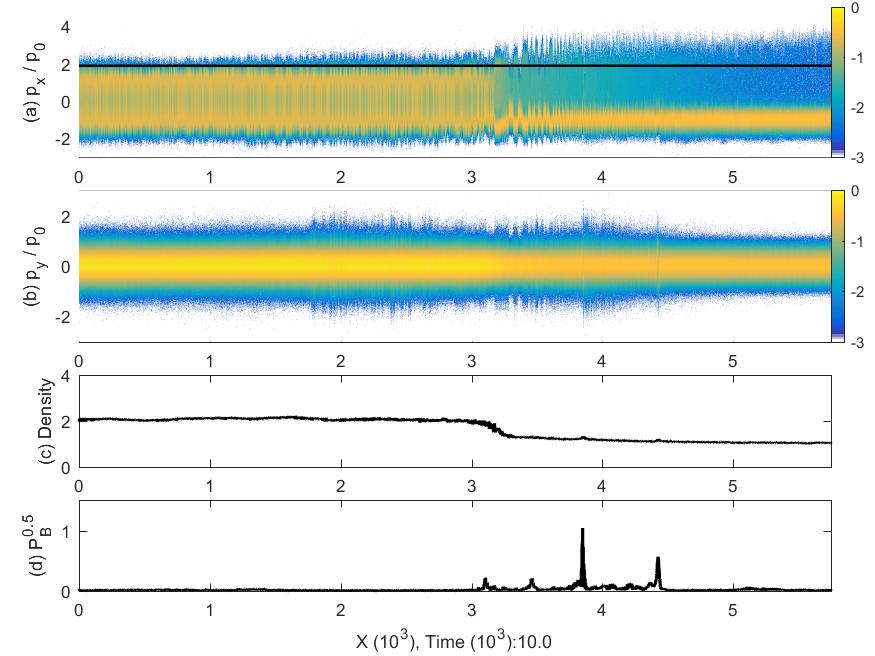}
	\caption{$\omega_c =0.8$}
    \label{w0.8-1d-phase-space}
    \end{subfigure}
    
	\caption{1D simulation results at $t=10^4$ for different values of $\omega_c$. Panels (a) and (b) show the phase space log-density distributions $\log_{10}f(x,p_x)$ and $\log_{10}f(x,p_y)$, normalized to $p_0$, the peak value at $t=0$. Panel (c) shows the total number density in units of $n_0/2$ while (d) shows $P_B^{1/2}$. The black horizontal line in the top panel marks $p_x/p_0=m_ev_A$.}
    \label{merged-1d-phase-space}
\end{figure}

We next consider the steady state configuration of the shock formed. In \autoref{merged-1d-phase-space}, we show the phase space density distribution,  the density
and the square root of the normalized magnetic pressure $P_B = ({(B_x-B_0)}^2+B_y^2+B_z^2)/2\mu_0$  for the three cases considered here ($\omega_c = 0.2, 0.4$, and $0.8$) at time $t = 10^4$, where the generated shocks reach a (quasi-) steady state. At a stronger magnetic field, the Alfv\'en speed is higher, and hence the shock speed is higher and the shock propagates faster, as can be seen in the figure (panels (a) and (b)). 

At lower magnetic fields, more Alfv\'enic waves are accumulated near the shock front, implying a sharper discontinuity of the fluid properties at the front (panels (c)) as well as stronger magnetic pressure (panels (d)). This amplification of the waves due to compression by the shock, as well as deceleration which is expected since the Alfvén speed is inversely proportional to the number density of charge carriers.

From the results in \autoref{merged-1d-phase-space}, one can discern distinct regimes of shock behavior as a function of the cyclotron frequency $\omega_c$. The shock formed in the presence of a weaker magnetic field, $\omega_c=0.2$ is slowest, its density highest, and the magnetic pressure is located primarily at the shock. With increasing background field strength one sees smoothing of the discontinuity as well as more pronounced anisotropy in momentum space. For the strongest magnetic field ($\omega_c=0.8$) no shock is formed, as can be seen by the factor 2 in the  density jump.

\begin{figure}
	\includegraphics[width=\columnwidth]{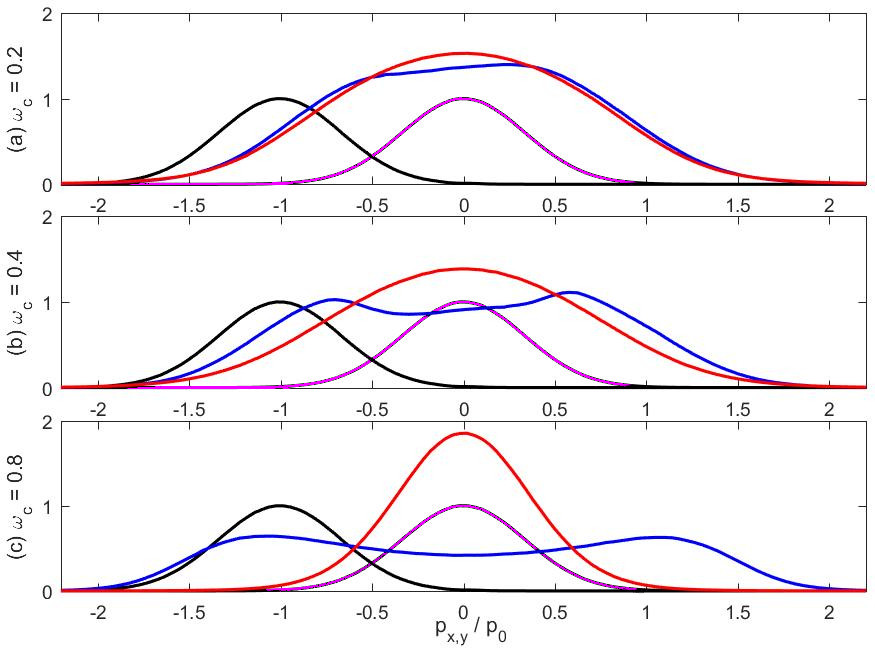}
	\caption{Momentum distributions averaged over $x<10^3$: Panel (a) shows
		that of the simulation with $\omega_c = 0.2$, (b) that of the simulation
		with $\omega_c = 0.4$ and (c) that of the simulation with $\omega_c = 0.8$.
		The black curve shows the initial plasma distribution along $p_x$ and the purple
		  corresponds to $p_y$.
        Blue and red curves show the distributions at $t=1336$ along $p_x$ and
		$p_y$. All curves are normalized to the peak value of the initial distribution. Due to the particle flow, the later curves contain more particles and therefore cover a larger area than the initial curves.}
	\label{momentum_distribution}
\end{figure}

In \autoref{momentum_distribution} we show the momentum distribution functions of the electrons and positrons in the region $0<x<10^3$, comparing the initial and intermediate ($t=1336$) states. From these one can see how the distributions, which are initially thermal with equal temperature in each direction, have been modified by interaction with the shock and the associated magnetic turbulence. We examine each of the three cases $\omega_c=0.2,0.4,0.8$ to assess the effect of each magnetic field strength on the timescale for relaxation to thermal equilibrium. Since anisotropic velocity distribution is the source of the Alfv\'enic turbulence this figure demonstrates the time evolution of the wave growth rate.

The simulation state in \autoref{momentum_distribution}(a) has reached approximate equilibrium. The increase in the enclosed area for the final curves represents a higher particle count in the considered region due to the compression of the plasma in the downstream. Apart from a small deviation in the $p_x$ distribution we observe the entropy-maximizing case of a thermal and isotropic distribution.

The state shown in \autoref{momentum_distribution}(b) is still evolving towards an equilibrium; the blue curve shows peaks in the incoming and reflected velocity direction but the blue peak at negative $p_x$ is at a lower speed than the initial beam mean speed. The perpendicular direction
has also been heated which is evidenced by the flatter distribution. 

\autoref{momentum_distribution}(c) is clearly far from equilibrium. The red curve is approximately double the initial distribution, because incoming and reflected beams have been added together, but similar same mean speed and temperature are seen along that direction. The peaks of the blue curve are located at $p_x \approx \pm m_ev_0$.

Particles that leak from the overlap layer into the foreshock are both hot and
anisotropic, leading to a firehose instability \cite{Pokhotelov06}. This is a result of the broad
distribution of $v_\parallel$. The waves in the $\omega_c=0.8$ case escape upstream
and away from the boundary overlap layer and foreshock.

Building on our previous conclusion \cite{Dieckmann2020} that $\omega_c = 0.1$ rapidly leads to
equilibration we find the following: the configuration with $\omega_c = 0.2$ likewise converges rapidly, $\omega_c = 0.4$ does so more slowly but notably in our results, and  $\omega_c = 0.8$ does not appear to reach an equilibrium within the time considered. These results provide an approximate answer to our initial research question regarding the effect of the background field on the formation and stability of the pair-Alfv\'en shock.

\subsection{2D Simulations}

In the previous section, we have shown in the 1D simulations that the largest magnetic field
suppresses the formation of a shock and excludes all wave modes not satisfying
$\mathbf{k} \parallel \mathbf{x}$. Here, we extend the analysis by considering 2D simulations of the same environment, in order to examine if the same suppression occurs. 

\begin{figure}
	\includegraphics[width=\columnwidth]{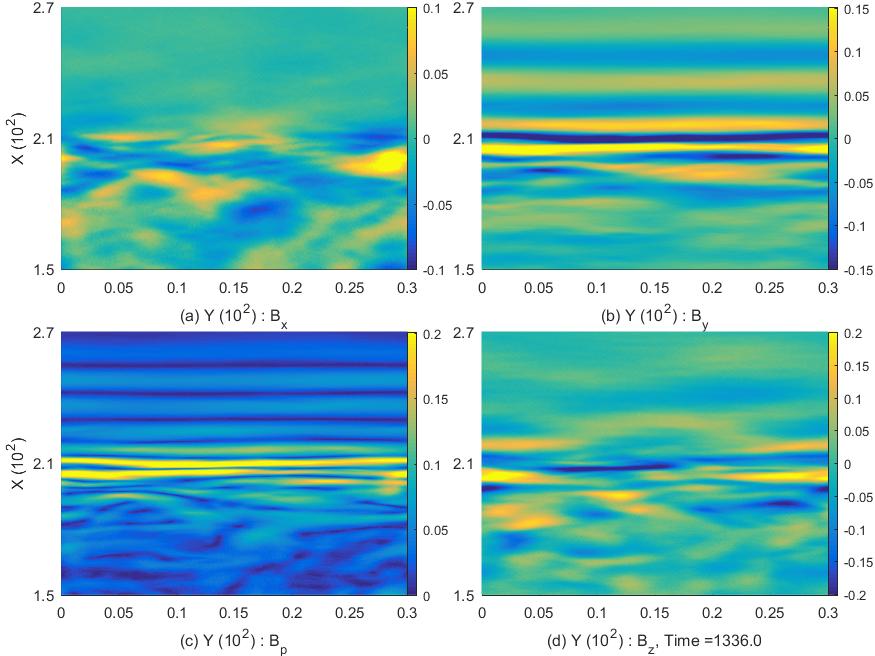}
	\caption{Magnetic field distribution near the shock for $\omega_c = 0.2$ and
		$t=1336$: Panel (a) shows the magnetic $B_x$ component. Panel (b) shows
		$B_y$. Panel (c) shows $B_p = {({(B_x-B_0)}^2+B_y^2)}^{1/2}$ and (d) shows
		$B_z$. In the normalized units we use here, the maximum value of the magnetic field is 0.2, similar to $B_0$.}
	\label{magneticCase2}
\end{figure}

In \autoref{magneticCase2} we plot the magnetic field
distribution near the shock at time $t=1336$, for the case $\omega_c = 0.2$. 
The amplitudes of the magnetic field components are comparable to that of $B_0$ and the waves are thus
well into the nonlinear regime. The pair-Alfvén wave is practically planar. As in the 1D case (panels(d) in \autoref{merged-1d-phase-space}) we can see a concentration of magnetic energy in the vicinity of the shock front, at $x\approx200$; see panel (c). Also similarly to the 1D case, the shock is much faster than the waves which results in the accumulation of magnetic energy at the shock front which is amplified as it crosses the discontinuity and is compressed.

\begin{figure}[h!]
\begin{subfigure}{0.475\textwidth}
	\includegraphics[width=\columnwidth]{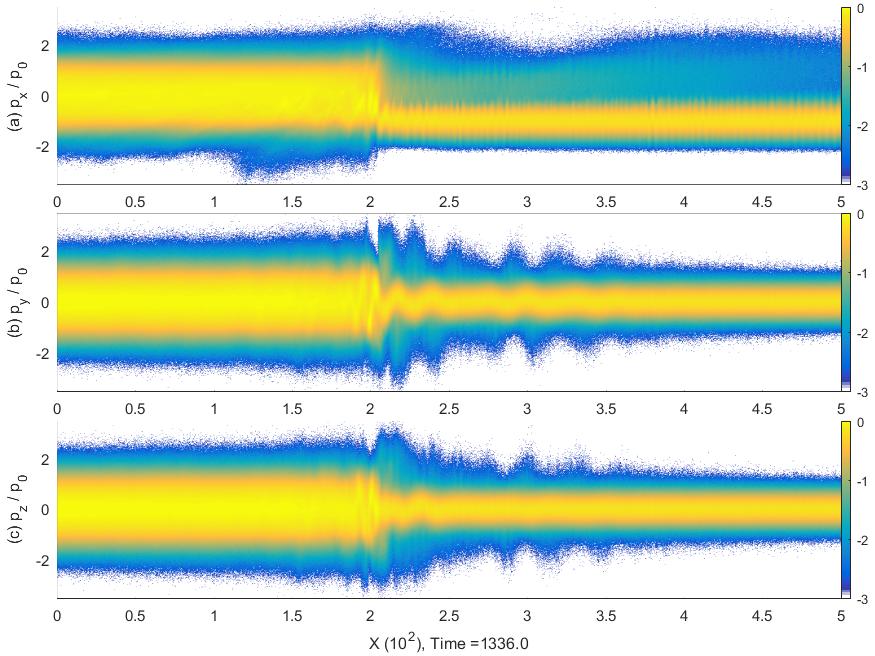}
	\caption{$\omega_c=0.2$ }
	\label{phaseCase2}
\end{subfigure}

\begin{subfigure}{0.475\textwidth}
	\includegraphics[width=\columnwidth]{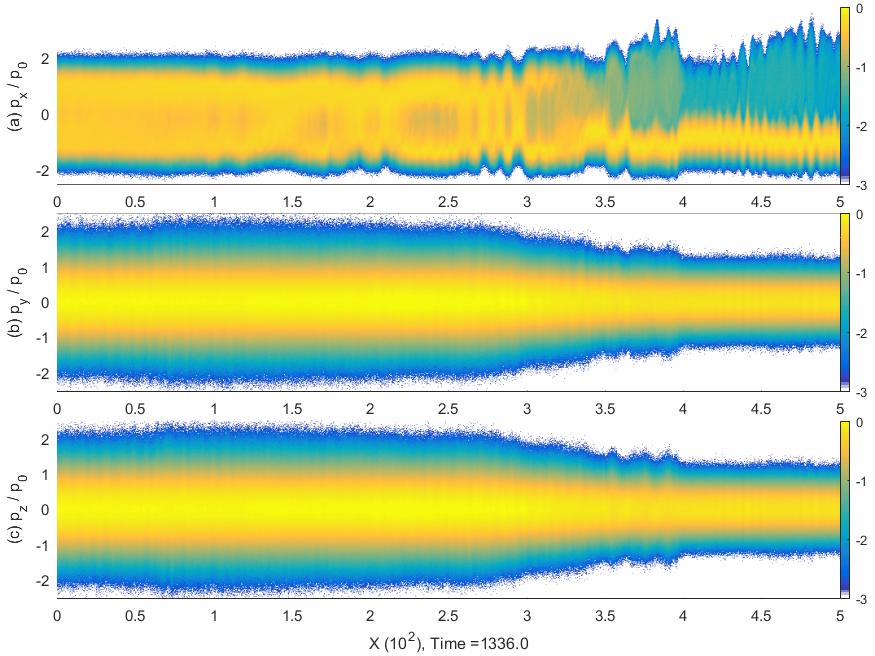}
	\caption{$\omega_c=0.8$ }
	\label{phaseCase8}
\end{subfigure}

\caption{Phase space density distribution at the time $t=1336$ for different values of $\omega_c$. 
    All distributions have been averaged over $y$.
	Panel (a) shows the projection of the distribution onto the $x,p_x$ plane.
	Panels (b) and (c) show the projections onto the two other momentum
	directions. Phase space densities are normalized to their initial peak value
	and displayed on a base-10 logarithmic scale.}
\label{case-phase-merged}
\end{figure}

The $y$-averaged phase space density distribution is shown in
\autoref{phaseCase2}, with downstream and upstream regions separated by a sharp transition layer. A
population of energetic particles is moving upstream, as is seen in  \autoref{phaseCase2}(a) [left side of the figure]. Their momenta perpendicular to the flow direction are relatively small, since they originate from a thermal distribution
and only those whose momentum lies mostly along $\mathbf{x}$ can outrun the shock. Particles shown in panels (b) and (c) move on helical
orbits under the influence of the magnetic field. They have a large wavelength
upstream which is diminished after crossing the shock due to the relatively
stronger background magnetic field. The shock jump conditions enforce frequency matching at the shock and thus a change in the Alfv\'en speed across the shock.

The results obtained with $\omega_c = 0.8$ (\autoref{phaseCase8}) show the same phase space density distribution as in \autoref{phaseCase2} but this time for the strong magnetic field case. As opposed to the results seen in \autoref{phaseCase2}, no shock appears in \autoref{phaseCase8}, as is evident from the fact that no sharp discontinuity in the particle momentum is observed.

The downstream plasma is cooler along $y$ and $z$ than along $x$ and the
distribution as seen in \autoref{phaseCase8}(a) is non-Maxwellian. The
shape of the distribution exhibits electrostatic phase space holes. This represents
a plasma transition layer that is not bounded by a shock.

\begin{figure}
	\includegraphics[width=\columnwidth]{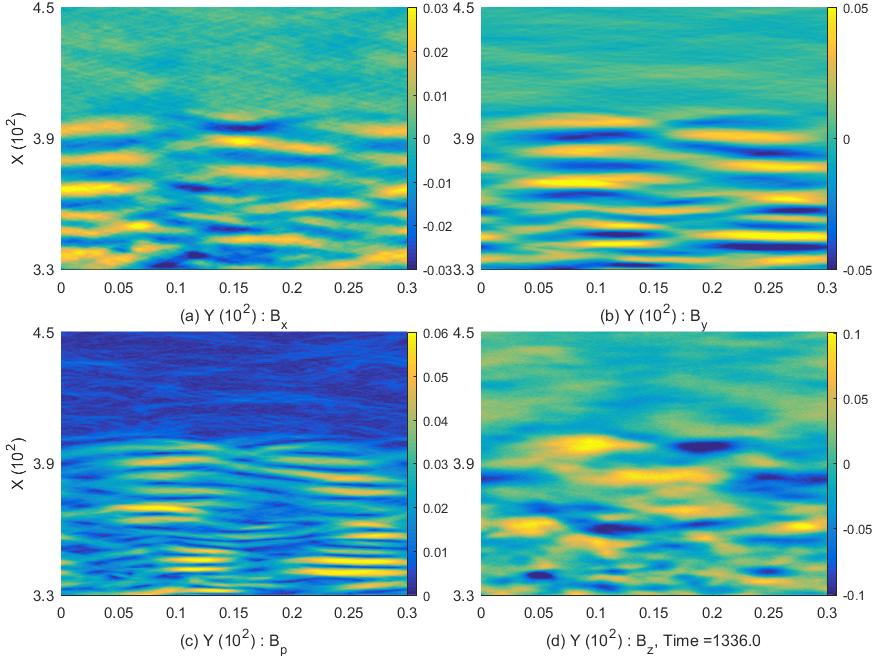}
	\caption{Magnetic field distribution near the shock for $\omega_c = 0.8$ and
		$t=1336$: Panel (a) shows the magnetic $B_x$ component. Panel (b) shows
		$B_y$. Panel (c) shows $B_p = {({(B_x-B_0)}^2+B_y^2)}^{1/2}$ and (d) shows
		$B_z$. Note that the in this case the magnitude of the magnetic field is much smaller than $B_0$.}
	\label{magneticCase8}
\end{figure}

The magnetic field distributions in \autoref{magneticCase8} show clearly distinguished waves in the $xy$ plane. In contrast to the $\omega_c=0.2$ case we see a smooth change in density along $x$ with no clear shock front. We also observe a periodic variation in y, implying a spontaneous generation of oblique waves. Although this variability was present in the lower $\omega_c$ cases, it was not pronounced because the presence of the shock wave implied that these relatively small fluctuations were dominated by waves oriented along the $x$ (shock propagation) direction. The box size along $y$ is large enough to resolve one wave oscillation along this direction.  

Having $k_y \neq 0$ implies the presence of spontaneously generated oblique waves (distinct from the 1D case where obliquity was imposed by rotating the background field). These waves have their speed reduced by a factor of $\cos\alpha$ and cannot outrun the expanding plasma, c.f. \autoref{oblique}. Their amplitude is much lower than that of the corresponding waves for the lower magnetization case shown in \autoref{magneticCase2} because the Alfv\'en waves are too fast to accumulate at the shock front. We also see fast magnetosonic waves but they are not well-structured.

We also repeated the simulation with a $y$-scale increased by a factor of $10$, to examine the possibility of spurious effects arising from the small size relative to the wavelength of these observed fluctuations. The results appear similar in the simulations with the original and larger boxes, in particular the wavelength was not affected. This indicates that although the computationally convenient assumption of a periodic boundary along $y$ is unphysical, it does not significantly affect the outcome. We do notice a slightly stronger $B_z$ but this might be coincidental and does not invalidate our conclusions.

The degree to which the Alfvén wave thermalizes the plasma determines the
speed of the shock. If significant energy is transferred into the perpendicular
degrees of freedom of the particles then the compression ratio will be higher
and hence the shock speed will be lower.

In contrast to the case discussed in \autoref{two-stream-and-firehose}, the overlap layer for the $\omega_c=0.8$ case is not firehose-unstable. Furthermore the $\omega_c=0.4$ case appears to lie close to the stability threshold.

\section{Discussion}

In this work, we studied the formation of Pair-Alfv\'en shocks in the presence of a range background magnetic field strengths. These discontinuities had previously been identified only in simulations with weak magnetization\cite{Dieckmann2019a}. Using both 1D and 2D simulations with the PIC code EPOCH, we identify the upper magnetization threshold beyond which these shocks cannot form, explain the suppression mechanism via anisotropy-inhibited Alfvén wave growth, and distinguish three distinct regimes of shock behavior based on magnetic field strength. It also confirms that oblique Alfvén modes, which cannot appear in 1D, do not compensate for this suppression. Hence we have provided new constraints on shock formation in such collisionless pair plasmas as are expected to be produced at e.g. the base of jets.

Our Particle-In-Cell simulations demonstrate that these recently discovered pair-Alfvén shocks are sustained by self-generated Alfvén waves that overtake the shock front in the upstream plasma. The simulations show that the shock only develops when the background magnetic field and associated Alfvén speed are sufficiently low to enable this wave amplification process. At higher magnetizations, our results indicate the Alfvén wave growth is suppressed and the increased speed means that waves outrun the shock, prohibiting the formation of a sustained pair-Alfvén shock. This aligns with theoretical expectations for the firehose instability threshold, which constrains the allowable thermal anisotropy at high plasma beta. The simulated cases with $\omega_c = 0.4$ and $0.8$ appear to lie near or beyond this stability limit, quenching the waves required to mediate the shock.

The plots in \autoref{merged-1d-phase-space} concisely exhibit the three regimes of instability growth (in the 1D case); weak, intermediate, and strong. The sharpness (or absence) of the shock front, concentration of magnetic energy, and momentum distribution of the charged particles are all clearly distinguished. As we explain, these differences are a result of the ability of the Alfv\'enic turbulence to outrun the shock, as well as the dependency of the relevant plasma instabilities on the difference between the momentum distribution functions in the directions parallel and perpendicular to the background field, or equivalently the shock propagation direction.

The detailed structure in 2D is shown in \autoref{magneticCase2} and \autoref{magneticCase8}, where the we see clear formation of a shock front with $x$-aligned normal in only the weak case. Contrarily the strong field case has relatively diffuse magnetic energy, of which a significant fraction occupies oblique modes. The relevant constraints on anisotropy \autoref{anisotropy-vs-alfven-speed} and \autoref{thermal-anisotropy-bound} are not satisfied here, and so the plasma is remains non-thermal, as shown in \autoref{momentum_distribution}.

The three cases considered in this work ($\omega_c=0.2,0.4,0.8$) of respectively weak, moderate, and strong background fields are ultimately distinguished by tendency to equilibrium. Specifically, the weak field case ($\omega_c = 0.2$) shows rapid evolution towards equilibrium, the moderate field case ($\omega_c = 0.4$) evolves more slowly, and the strong field case ($\omega_c = 0.8$) exhibits no clear evolution towards equilibrium. This resolves our questions raised in our previous work\cite{Dieckmann2020} where we had started by examining the formation of the pair-Alfvén shock in the weak case.

Notably, our 2D simulations reveal the presence of oblique Alfvén modes not captured in 1D. However, these oblique waves are still suppressed at high magnetizations and do not lead to shock growth. The oblique modes propagate too slowly to outrun the expanding downstream plasma for the strong field case.
The absence of the shock was noted in \cite{Bret2017}. We can now explain this since the magnetization is outside the allowed range we identified.

Since these shocks are potentially found across a wide variety of high-energy astrophysical environments, our result is potentially relevant to observational studies of shock emission in high-energy environments. In particular we have provided an upper limit on the strength of the magnetic field above which no shock will form. If a pair-Alfv\'en shock can be identified observationally, one could then conclude an upper limit on the magnetization, $\omega_c \leq 0.8$ which can constrain such predictions as models of jet launching.

We defer to future work the question of whether 3D PIC simulation studies will show whether pair-Alfvén wave could replace the filamentation mode in a realistic geometry. Furthermore, 3D simulations can determine whether pair-Alfvén waves keep their linear polarization in the nonlinear regime. If such shock are to be considered as acceleration sites for cosmic rays then a population of heavy particles (e.g. protons) can be added to the simulations to see if they are efficiently accelerated, and if the presence of these particles and their draw of energy from the shock affects its formation and structure.

\begin{acknowledgments}
	The authors also wish to acknowledge the DJEI/DES/SFI/HEA Irish Centre for
	High-End Computing (ICHEC) for the provision of computational facilities.
	Derived data supporting the findings of this study are available from the
	corresponding author upon reasonable request.
\end{acknowledgments}

\nocite{*} \bibliography{three,refs}
\bibliographystyle{aasjournalv7}

\end{document}